\documentclass[12pt]{iopart}

\usepackage{iopams}
\usepackage{braket}
\usepackage{amsfonts}
\usepackage{color}
\usepackage[dvipsnames]{xcolor}
\usepackage{amssymb}
\usepackage{mathrsfs}
\usepackage{graphicx}
\usepackage{lmodern}
\usepackage{lineno}
\usepackage{hyperref}
\usepackage[normalem]{ulem}
\usepackage{soul}

\usepackage{lipsum}

\newcommand{\beq}{\begin{eqnarray}}
\newcommand{\eeq}{\end{eqnarray}}

\newcommand{\Exp}{\mathrm{Exp}_{\star}}
\newcommand{\nn}{\nonumber}

\def\keywords#1{\vspace{10pt}
     \begin{indented}
     \item[]\rm Keywords: #1\par
     \end{indented}}

\begin{document}



\title{Star exponentials and Wigner functions for time-dependent harmonic oscillators}
\author{J.~Berra--Montiel$^{1,2}$, D.~Contreras--Bear$^{1}$, A.~Molgado$^1$ and 
M.~S\'anchez--C\'ordova$^{1}$}

\address{$^{1}$ Facultad de Ciencias, Universidad Aut\'onoma de San Luis 
Potos\'{\i} \\
Campus Pedregal, Av. Parque Chapultepec 1610, Col. Privadas del Pedregal, San
Luis Potos\'{\i}, SLP, 78217, Mexico}
\address{$^2$ Dipartimento di Fisica ``Ettore Pancini", Universit\'a degli studi di Napoli ``Federico II", Complesso Univ. Monte S. Angelo, I-80126 Napoli, Italy}

\eads{\mailto{\textcolor{blue}{jasel.berra@uaslp.mx}},\
\mailto{\textcolor{blue}{a324797@alumnos.uaslp.mx}},\
\mailto{\textcolor{blue}{alberto.molgado@uaslp.mx}},\
\mailto{\textcolor{blue}{maria.cordova@uaslp.mx}} 
}


\begin{abstract}

In this paper, we address the Wigner distribution and the star exponential function for a time-dependent harmonic oscillator for which the mass and the frequency terms are considered explicitly depending on time.     To such an end, we explore the connection between the star exponential, naturally emerging within the context of deformation quantization, and the propagators constructed through the path integral formalism.   In particular, the Fourier-Dirichlet expansion of the star exponential implies a distinctive quantization of the Lewis-Riesenfeld invariant.  Further, by introducing a judicious time variable, we recovered a time-dependent phase function associated with the Lewis-Riesenfeld construction of the standard Schrödinger picture.  In particular, we applied our results to the cases of the Caldirola-Kanai and the time-dependent frequency harmonic oscillators, recovering relevant results previously reported in the literature.

\end{abstract}



\section{Introduction}

The ubiquitous time-dependent harmonic oscillator is a fundamental model in both classical and quantum mechanics due to its broad range of applications and inherent mathematical elegance. Unlike its time-independent counterpart, the time-dependent harmonic oscillator incorporates time-varying parameters, such as the mass and the frequency, which introduce additional complexity, making it a rich subject for theoretical investigation. Exact solutions for this system has been extensively studied from different analytical perspectives.  
In particular, it finds applications in diverse fields such as Quantum Optics \cite{Colegrave, Gardiner}, Plasma Physics \cite{Lo}, Quantum Field Theory~\cite{Vergel, Birrell}, String theory~\cite{Papado, Nardi}, Cosmology \cite{Berger, Dantas, Barbero}, Soliton theory~\cite{Gromov, Hdz}, and Nanotechnology \cite{Brown}, to mention some areas where the time-dependent harmonic oscillator commonly emerges in phenomena associated with either oscillatory damping or time-varying potentials. A well-known example of such situation is given by the Caldirola-Kanai oscillator, a notable variant of the time-dependent harmonic oscillator that models dissipative systems and provides insights into quantum states in non-static and fluctuating backgrounds \cite{Caldirola, Kanai, Razavy, Unruh}.

From a methodological point of view, the time-dependent harmonic oscillator model has inspired the development of different techniques such as the Lewis-Riesenfeld invariant operator method \cite{LR, Choi}. This method involves constructing an invariant operator and expressing Schr\"odinger’s wave functions in terms of invariant eigenstates, supplemented by an explicitly time-dependent phase. However, constructing an invariant operator can be quite difficult in some cases, as it often requires the use of canonical transformations via unitary operators, a process that can become highly intricate in practice due to the appearance of nonlinearities.  

In consequence, the main aim of this paper is to address the time-dependent harmonic oscillator from the perspective of the deformation quantization formalism.   Such a formalism stands for a quantization in phase space in which the algebraic structures in Classical Mechanics are to be deformed into an associative non-commutative product, namely the star product,  that embodies the Quantum Mechanical properties of a given system.  In particular, within the deformation quantization program, the time evolution is encoded by 
the star exponential of the Hamiltonian function which, as recently exposed in~\cite{starexpo}, may be constructed employing the quantum mechanical propagator. For the time-dependent harmonic oscillator, we can obtain the star exponential in terms of the Lewis-Resienfeld invariant by introducing an appropriate non-linear time variable.   Hence, the Lewis-Riesenfeld invariant becomes quantized and we straightforwardly recover a time-dependent phase function commonly obtained within the Schrödinger scheme by means of eigenfunctions associated with such an invariant term.  As a byproduct, we are able to successfully analyze the Caldirola-Kanai and the time-dependent frequency harmonic oscillators, recovering previously reported results obtained by some other quantum mechanical techniques.

The rest of the paper is as follows. In Section~\ref{sec:DQ-Star}, we briefly review the deformation quantization program, emphasizing in the Moyal product, the star exponential and, particularly, in the relation of the latter with the quantum mechanical propagator.  
In Section~\ref{sec:star-TDHO}, we define the Lagrangian for our model and introduce an appropriate non-linear time variable for which the star exponential and the Wigner distribution may be explicitly constructed in terms of the Lewis-Riesenfeld invariant.
Sections~\ref{sec:CK} and~\ref{sec:TD-Freq} are devoted to explore our results 
for the Caldirola-Kanai and time-dependent frequency cases.   Finally, in Section~\ref{sec:conclu} we include some remarks and perspectives.

\section{Deformation quantization and the star exponential}
\label{sec:DQ-Star}
The origins of the deformation quantization formalism can be traced back to the Weyl quantization map \cite{Wigner} which provides a correspondence between classical observables and quantum operators. More specifically, for any classical observable $f(\mathbf{x},\mathbf{p})\in L^{2}(\mathbb{R}^{2n})$, defined on the phase space $\Gamma=\mathbb{R}^{2n}$, the Weyl quantization map associates this observable with a quantum operator $\mathcal{Q}_{W}(f)$ acting on the Hilbert space $\mathcal{H}=L^{2}(\mathbb{R}^{n})$ through the following relation
\begin{equation} \label{Weyl} 
\mathcal{Q}_{W}(f) = \frac{1}{(2\pi)^n} \int_{\mathbb{R}^{2n}} \widetilde{f}(\mathbf{a}, \mathbf{b}) e^{i(\mathbf{a} \cdot \widehat{\mathbf{X}} + \mathbf{b} \cdot \widehat{\mathbf{P}})} \,d\mathbf{a}\, d\mathbf{b}, \end{equation}
where $\widetilde{f}(\mathbf{a},\mathbf{b})$ denotes the Fourier transform of $f(\mathbf{x},\mathbf{p})$ over $\mathbb{R}^{2n}$, with $\mathbf{a},\mathbf{b}\in\mathbb{R}^{n}$. The components of the operators $\widehat{\mathbf{X}}$ and $\widehat{\mathbf{P}}$ are subject to the canonical commutation relations, represented by $[\widehat{X}_{i},\widehat{P}_{j}]=i\hbar\delta_{ij}$, where $i,j=1,\ldots, n$. Rigorously, the integral~(\ref{Weyl}) is computed within the weak operator topology and may not exhibit absolute convergence \cite{Takhtajan} (see also \cite{Dias}, and references therein).
Since the Weyl transform involves an integral of an operator over the phase space, its integral kernel can be computed as
\begin{equation}
\kappa_{f}(\mathbf{x},\mathbf{x}')=\braket{\mathbf{x}|\mathcal{Q}_{W}(f)|\mathbf{x}'}.
\end{equation}Using the Baker-Campbell-Hausdorff formula, we have
\begin{equation}
\kappa_{f}(\mathbf{x},\mathbf{x}')=\frac{1}{(2\pi\hbar)^{n}}\int_{\mathbb{R}^{n}}f\left(\frac{\mathbf{x}+\mathbf{x}'}{2},\mathbf{p} \right)e^{i\mathbf{p}\cdot(\mathbf{x-\mathbf{x'}})/\hbar}d\mathbf{p}. 
\end{equation}
Additionally, the integral kernel fulfills the following relation
\begin{equation}
\int_{\mathbb{R}^{2n}}|\kappa_{f}(\mathbf{x},\mathbf{x}')|^{2}d\mathbf{x}d\mathbf{x}'=\frac{1}{(2\pi\hbar)^{n}}\int_{\mathbb{R}^{2n}}|f(\mathbf{x},\mathbf{p})|^{2}d\mathbf{x}d\mathbf{p}.
\end{equation}
This implies that the operator $\mathcal{Q}_{W}(f)$ belongs to the Hilbert-Schmidt class of operators \cite{Reed}, which guarantees the existence of a well-defined (though possibly infinite) trace \cite{Folland}. Furthermore, the Weyl quantization maps functions from $L^{2}(\mathbb{R}^{n})$ into functions in $L^{2}(\mathbb{R}^{n})$ through the relation 
\begin{equation}
\mathcal{Q}_{W}(f)\psi(\mathbf{x})=\int_{\mathbb{R}^{n}}\kappa_{f}(\mathbf{x},\mathbf{x}')\psi(\mathbf{x'})d\mathbf{x}',
\end{equation}
for any $\psi\in L(\mathbb{R}^{n})$. As a result, the Weyl transform (\ref{Weyl}) provides a mapping from the space of square-integrable functions defined on the phase space $L^{2}(\mathbb{R}^{2n})$ to the space of Hilbert-Schmidt operators $\mathrm{HS}(L^2(\mathbb{R}^n))$ acting on the Hilbert space $L^2(\mathbb{R}^n)$. For real-valued functions, the corresponding Weyl transform results in a self-adjoint operator \cite{Curtright}. An explicit inverse map of the Weyl transform, $\mathcal{Q}_{W}^{-1}:\mathrm{HS}(L^2(\mathbb{R}^n))\to L^{2}(\mathbb{R}^{2n})$, can be determined as follows 
\begin{equation}\label{inverseWeyl}
\mathcal{Q}_{W}^{-1}(\hat{A})=A_{W}(\mathbf{x},\mathbf{p})=\int_{\mathbb{R}^{n}}\braket{\mathbf{x}-\frac{\mathbf{x}'}{2}|\hat{A}|\mathbf{x}+\frac{\mathbf{x}'}{2}}e^{i\mathbf{p}\cdot\mathbf{x}'/\hbar}d\mathbf{x}',
\end{equation}for any operator $\hat{A}\in \mathrm{HS}(L^2(\mathbb{R}^n))$. This function is called the Wigner transform of the operator $\hat{A}$ \cite{Takhtajan}. Following the standard terminology in harmonic analysis \cite{Reed,Folland}, we will say that the Weyl inversion formula (\ref{inverseWeyl}) defines a symbol, denoted by $A_{W}(\mathbf{x},\mathbf{p})$, from it quantization $\hat{A}$. Even though it is possible to generalize Weyl quantization to the space of generalized functions, for the sake of simplicity, we restrict our attention to integrable functions. We refer the reader to \cite{Folland} for a detailed treatment of this generalization.

With the Weyl quantization map and its inverse at hand, we are able to define the star product. Since Weyl's map  $\mathcal{Q}_{W}: L^{2}(\mathbb{R}^{2n}) \to \mathrm{HS}(L^{2}(\mathbb{R}^{n}))$ is bijective and the product of Hilbert-Schmidt operators is closed \cite{Folland}, it follows that there exists a unique function in $L^{2}(\mathbb{R}^{2n})$, denoted by $f \star g$, such that
\begin{equation}
\mathcal{Q}_{W}(f)\mathcal{Q}_{W}(g)=\mathcal{Q}_{W}(f\star g)  \,.
\end{equation} 
By applying the Wigner transform (\ref{inverseWeyl}) on both sides, we have
\begin{equation}\label{starWW}
(f\star g)(\mathbf{x},\mathbf{p})=\mathcal{Q}_{W}^{-1}(\mathcal{Q}_{W}(f)\mathcal{Q}_{W}(g)) \,.
\end{equation}
This implies that the symbol corresponding to the product of two operators is a $\hbar$-dependent function on phase space, known as the star product, which can be explicitly written as
\begin{equation}\label{integralstar}
\hspace{-8ex}
(f\star g)(\mathbf{x},\mathbf{p})=\frac{1}{(\pi\hbar)^{2}}\int_{\mathbb{R}^{4}}f(\mathbf{a},\mathbf{b})g(\mathbf{c},\mathbf{d})e^{-\frac{2i}{\hbar}(\mathbf{p}\cdot(\mathbf{a}-\mathbf{c})+\mathbf{x}\cdot(\mathbf{d}-\mathbf{b})+(\mathbf{c}\cdot\mathbf{b}-\mathbf{a}\cdot{d}))}d\mathbf{a}\,d\mathbf{b}\,d\mathbf{c}\,d\mathbf{d} \,.
\end{equation}  
This expression is known as the integral representation of the star product, and in the limit $\hbar \to 0$ it reduces to the pointwise product \cite{Folland}
\begin{equation}
\lim_{\hbar\to 0}f\star g=fg.
\end{equation}
Moreover, for $f,g\in C^{\infty}(\mathbb{R}^{2n})$, both functions can be Taylor-expanded within the integral in (\ref{integralstar}), resulting in the differential representation of the star product \cite{Teaching},
\begin{eqnarray}\label{Star}
(f\star g)(\mathbf{x},\mathbf{p})&=&\sum_{m,n=0}^{\infty}\left( \frac{i\hbar}{2}\right)^{m+n}\frac{(-1)^{m}}{m!n!}\left( \partial^{m}_{\mathbf{p}}\cdot\partial^{n}_{\mathbf{x}}f\right)\left( \partial^{n}_{\mathbf{p}}\cdot\partial^{m}_{\mathbf{x}}g\right) \nonumber \\  
&=&f(\mathbf{x},\mathbf{p})\exp\left[\frac{i\hbar}{2}\left(\frac{\overleftarrow{\partial}}{\partial \mathbf{x}}\cdot\frac{\overrightarrow{\partial}}{\partial\mathbf{p}}- \frac{\overleftarrow{\partial}}{\partial \mathbf{p}}\cdot\frac{\overrightarrow{\partial}}{\partial\mathbf{x}}\right)\right]g(\mathbf{x},\mathbf{p})   \,,  
\end{eqnarray}   
where in the last line the differential operators $\overleftarrow{\partial}/\partial \mathbf{x}$  and $\overleftarrow{\partial}/\partial \mathbf{p}$ act on the left, while the differential operators $\overrightarrow{\partial}/\partial \mathbf{x}$ and $\overrightarrow{\partial}/\partial \mathbf{p}$ act on the right. In other words, as pointed out in \cite{Bayen}, this product represents an associative deformation of the standard commutative pointwise product of functions on $C^{\infty}(\mathbb{R}^{2n})$.
Furthermore, by defining the Moyal bracket as
\begin{equation}
\{f,g\}_{M}=\frac{1}{i\hbar}\left(f\star g-g\star f\right),    
\end{equation}
we may note that it is this bracket, and not the Poisson bracket, that corresponds to the quantum commutator, that is
\begin{equation}
\left\lbrace f,g \right\rbrace_{M}=\frac{1}{i\hbar}\mathcal{Q}^{-1}_{W}\bigg(\left[\mathcal{Q}_{W}(f),\mathcal{Q}_{W}(g) \right]  \bigg). 
\end{equation}
The expansion of the Moyal bracket in terms of the parameter $\hbar$ reveals a formal deformation of the underlying Poisson structure, which can be interpreted as a quantum correction \cite{Kontsevich}.

Within the deformation quantization approach, the von Neumann dynamical equation is translated, by means of the Weyl transform, into the Moyal equation of motion \cite{Curtright}, namely
\begin{equation}\label{Moyal}
\frac{\partial W(\mathbf{x},\mathbf{p})}{\partial t}=-\frac{i}{\hbar}\left( H\star W-W\star H\right)=\left\lbrace W,H \right\rbrace_{M}  \,,  
\end{equation}
where $H(\mathbf{x},\mathbf{p})$ and $W(\mathbf{x,\mathbf{p}})$ denotes the symbols corresponding to the quantum Hamiltonian operator and the density matrix normalized in the phase space, respectively. The function $W(\mathbf{x,\mathbf{p}})$, commonly known as the Wigner function \cite{Wigner}, is often referred to as a quasiprobability distribution. This designation arises from its capacity to characterize the probability distributions of quantum states in phase space. However, unlike classical probability distributions, the Wigner function can take on negative values in certain regions of phase space, reflecting the inherently non-classical and quantum mechanical nature of the system. These negative values are typically interpreted as signatures of quantum interference or non-classicality \cite{Generating, Recent}. In a sense, the Moyal equation generalizes Liouville's theorem from Classical Mechanics, as taking the limit $\hbar\to 0$ recovers, to some extent, the classical behavior of the system. The solution to Moyal dynamical equation (\ref{Moyal}) is formally expressed using the star exponential \cite{Bayen},
\begin{equation}\label{Exp}
\Exp\left(-\frac{i}{\hbar}tH \right)\equiv \sum_{n=0}^{\infty}\frac{1}{n!}\left(-\frac{i}{\hbar}t \right)^{n}H^{*n}   \,,  
\end{equation}  	
with $H^{*n}=H\star H\cdots\star H$ ($n$ factors) as
\begin{equation}
W(t)=\Exp\left(-\frac{i}{\hbar}tH \right)\star W\star \Exp\left(\frac{i}{\hbar}tH \right)   \,,
\end{equation}
where $W(t)$ denotes the Wigner function evolved in time from an initial  moment $t_0=0$ to a final one at $t$. Whenever $H$ corresponds to the Hamiltonian of the system, the star exponential admits a representation in terms of a  Fourier-Dirichlet expansion \cite{Bayen}, given by
\begin{equation}\label{FDexpansion}
\Exp\left(-\frac{i}{\hbar}tH \right)=2\pi\hbar\sum_{n=0}^{\infty}e^{-\frac{i}{\hbar}tE_{n}}W_{n}  \,,
\end{equation}
where $E_{n}$ denotes the eigenvalues of the Hamiltonian operator $\widehat{H}=\mathcal{Q}_{W}(H)$, and the phase space functions $W_{n}$ represent the diagonal components of the Wigner function, such that
\begin{eqnarray}
H\star W_{n} & = & E_{n}W_{n}   \,, \nn\\ 
W_{m}\star W_{n} & = & \delta_{mn}W_{n}  \,.
\end{eqnarray}
Moreover, for a continuous spectrum, the summations in the preceding expressions are replaced by integrals over a continuous variable representing energy, while the Kronecker delta functions are substituted with Dirac delta distributions \cite{Curtright}.

In general, the star exponential function (\ref{Exp}) belongs to the space of formal power series in $\hbar$ with coefficients in $C^{\infty}(\Gamma)$. For many examples this series converges in the distributional sense \cite{Bayen}. However, in most physically relevant cases, investigating the convergence of star products can present significant challenges due to the complex analytic structure of the formal series involved. In order to overcome such difficulties, it has been recently shown that the star exponential of a Hamiltonian function can be determined in terms of quantum mechanical propagators \cite{starexpo} by means of the explicit relation
\begin{equation}\label{StarE}
\Exp\left(-\frac{i}{\hbar}tH(q,p)\right)=2\int_{\mathbb{R}}e^{-\frac{i}{\hbar}2q'p}K(q+q',t,q-q',0)\, dq' \,,
\end{equation}
where $K(q,t,q',0)$ denotes the quantum mechanical propagator, and the square of its modulus, $|K(q,t,q',0)|^{2}$, represents the conditional probability distribution for finding a quantum particle at position $q\in\mathbb{R}$ at time $t$, given that it was at position $q'\in\mathbb{R}$ at time $t=0$. Equation (\ref{StarE}) offers an advantageous approach for calculating the star exponential function without relying on the convergence of formal series. This is achieved by using the Feynman's representation of the propagator in terms of path integrals, that is, expressing the quantum propagator as a weighted sum over all possible histories of the classical motion in the phase space. This perspective make certain aspects of quantum mechanics become more transparent as, for example, the fact that probability amplitudes of quadratic Lagrangians involve the classical action, or simply by introducing a convenient alternative to extend the analysis to the case of interacting systems and field theories \cite{starexpo, Sharan}. In the next sections we will investigate the properties of the star exponential function for the case of a time-dependent harmonic oscillator in connection with the path integral formalism. 

\section{Star exponential and Wigner functions for the Time-Dependent Harmonic Oscillator}
\label{sec:star-TDHO}

We begin our discussion with the classical Lagrangian for a time-dependent harmonic oscillator, given by
\begin{equation}\label{Lag}
    L(x,\dot{x}, t)=\frac{1}{2}m(t)(\dot{x}^2-\omega^2(t)x^2),
\end{equation}
where $m(t)$ and $\omega(t)$ are the time-dependent mass and frequency, respectively, while the dot stands for time-derivative. By applying the Euler-Lagrange equation to the Lagrangian in equation~(\ref{Lag}), the resulting equation of motion for the dynamical variable $x(t)$ reads
\begin{equation}\label{eomtdho}
    \ddot{x}+\gamma(t)\dot{x}+\omega^{2}(t)x=0,
\end{equation}
where whe have defined
\begin{equation}\label{motion}
    \gamma(t)=\frac{\dot{m}(t)}{m(t)}.
\end{equation}
Equation~(\ref{eomtdho}) describes the motion of an harmonic oscillator with a time-dependent frequency, influenced by a frictional force characterized by the coefficient $\gamma(t)$~(\ref{motion}). 

The canonical momentum associated with the Lagrangian (\ref{Lag}) is expressed as
\begin{equation}
    p=\frac{\partial}{\partial \dot{x}}L(x,\dot{x},t)=m(t)\dot{x},
\end{equation}
and the corresponding Hamiltonian is
\begin{equation}\label{Ham}
    H(x, p, t)= p\dot{x}-L=\frac{1}{2m(t)}p^2+\frac{1}{2}m(t)\omega^2(t)x^2.
\end{equation}
Let us consider now a point transformation $x\to\xi(x,t)=x/\rho(t)$, which represents a time-dependent scaling transformation with the scaling function $\rho(t)$ satisfying the auxiliary inhomogeneous equation
\begin{equation}\label{rho}
    \ddot{\rho}+\gamma(t)\dot{\rho}+\omega^2(t)\rho=\frac{\Omega^2}{m^2(t)\rho^3},
\end{equation}
for $\Omega$ constant. Without loss of generality, we will assume $\rho(t)>0$ \cite{Okada}. Under this point transformation, we have:
\begin{equation}\label{Lag2}
    L(x,\dot{x},t)=L(\xi, \dot{\xi},t)+\frac{d}{dt}\left(\frac{1}{2}m(t)\rho\dot{\rho}\xi^{2}\right),
\end{equation}
where
\beq
    L(\xi,\dot{\xi},t) 
& = &
\frac{m(t)}{2}\left[\rho^2\dot{\xi}^{2}-\rho(\ddot{\rho}+\gamma(t)\dot{\rho}+\omega^{2}(t)\rho)\xi^{2}\right]  \nn\\
& = &
\frac{m(t)}{2}\left[\rho^2\dot{\xi}^{2}-\left(\frac{\Omega^2}{m^2(t)\rho^2}\right)\xi^2
\right]  \,,
\eeq
where the last line was obtained in view of~(\ref{rho}).   As it is well-knwon, the total derivative term in~(\ref{Lag2}) does not affect the classical equation of motion; however, it plays a fundamental role as a generating function for canonical transformations. Further, in quantum mechanics, such a total derivative term manifests as a nontrivial time-dependent phase of the wave function \cite{Dirac}.
Now, by transforming the time variable  $t\to \tau(t)$, the Lagrangian (\ref{Lag2}) is mapped to
\begin{equation}\label{Lag3}
    L\left(\xi,\frac{d\xi}{d\tau},\tau \right)=\frac{m(\tau)}{2\dot{\tau}}\left[ \rho^{2}\dot{\tau}^{2}\left(\frac{d\xi}{d\tau}\right)^{2}-\left(\frac{\Omega^2}{m^2(\tau)\rho^2}\right)\xi^2 \right]\,,
\end{equation}
where the global parameter $\dot{\tau}$ in the denominator emerges to compensate the change of measure in the action induced by the transformation $t\rightarrow \tau$, and both $\rho$ and $\xi$ in this expression must be thought of as functions of the parameter $\tau$.   In general, the new time variable $\tau(t)$ may be considered as a nonlinear function in $t$. In particular, we can redefine the time variable $\tau(t)$ such that
\begin{equation}\label{dtau}
    \dot{\tau}=\frac{1}{m(t)\rho^{2}(t)} \,.
\end{equation}
This transformation implies that the coefficient of the kinetic energy term in equation (\ref{Lag3}) becomes constant. Consequently, we obtain

\begin{equation}\label{LHO}   
L\left(\xi, \frac{d\xi}{d\tau}, \tau\right)=\frac{1}{2}\left(\left(\frac{d\xi}{d\tau}\right)^{2}-\Omega^2\xi^2\right) \,.
\end{equation}
This Lagrangian describes the motion of a simple harmonic oscillator with constant frequency $\Omega$. The Euler-Lagrange equation for the dynamical variable $\xi(\tau)$ is given by
\begin{equation} \label{eomHO}
    \frac{d^{2}\xi}{d\tau^{2}}+\Omega^{2}\xi=0 \,.
\end{equation}
By writing the equation (\ref{eomHO}) in terms of the original dynamical variable $x$, we recover the equation of motion for a time-dependent harmonic oscillator, as given in (\ref{eomtdho}).
Furthermore, the canonical momentum $\pi$, conjugate to $\xi$, is defined as
\begin{eqnarray}
    \pi&=&\frac{\partial}{\partial(d\xi/d\tau)}L\left(\xi,\frac{d\xi}{d\tau},\tau\right)=\frac{d\xi}{d\tau}\nonumber\\
    &=&m(t)(\dot{x}\rho-x\dot{\rho}),
\end{eqnarray}
and the Hamiltonian of the system takes the form
\begin{eqnarray}\label{I}
    H(\xi,\pi)&=&\frac{1}{2}(\pi^{2}+\Omega^{2}\xi^{2}) \nonumber \\
    &=& \frac{1}{2}(p\rho -m(t)x\dot{\rho})^{2}+\Omega^{2}\frac{x^{2}}{\rho^{2}}\equiv I.
\end{eqnarray}
The quantity $I$ defined in this Hamiltonian is identified as the Lewis-Riesenfeld invariant which satisfies $dI/dt=0$ \cite{Pedrosa}. Relation~(\ref{I}) therefore implies that the Lewis-Riesenfeld invariant corresponds to the conserved energy in the $\xi$-coordinate system.

We now turn to the calculation of the star exponential function for the time-dependent harmonic oscillator. To achieve this, the quantum mechanical propagator of the Hamiltonian (\ref{I}) must first be determined. In the $\xi$-coordinate system, the Hamiltonian corresponds to a simple harmonic oscillator, thus, the propagator is given by~\cite{Takhtajan}
\begin{equation}
\hspace{-9ex} K(\xi_{2},\tau;\xi_{1},0) = \sqrt{\frac{\Omega}{2\pi i\hbar\sin{(\Omega\tau)}}} \exp\left[\frac{i m\Omega}{2\hbar\sin(\Omega\tau)}\left((\xi_{1}^{2}+\xi_{2}^{2})\cos(\Omega\tau)-2\xi_{1}\xi_{2}\right)\right] .       
\end{equation}
From (\ref{StarE}), we can read off that~\cite{starexpo}
\begin{eqnarray}\label{StarEho}
     \Exp{\left(-\frac{i}{\hbar}\tau H(\xi,\pi)\right)}&=&2\int_{\mathbb{R}}e^{-\frac{i}{\hbar}2\xi'\pi}K(\xi+\xi',\tau,\xi-\xi',0)d\xi' \nonumber\\
     &=& \left(\cos{\left(\frac{\Omega\tau}{2}\right)}\right)^{-1}\exp\left[\frac{2H(\xi,\pi)}{i\hbar\Omega}\tan{\left(\frac{\Omega\tau}{2}\right)}\right].
    \end{eqnarray}
The preceding star exponential admits a Fourier-Dirichlet expansion, as detailed in \cite{Bayen},
\begin{equation}\label{FD}
    \left(\cos{\left(\frac{\Omega\tau}{2}\right)}\right)^{-1}\exp\left[\frac{2H(\xi,\pi)}{i\hbar\Omega}\tan{\left(\frac{\Omega\tau}{2}\right)}\right]=\sum_{n}^{\infty}W_{n}(\xi,\pi)e^{-i\tau \Omega(n+1/2)},
\end{equation}
where the diagonal Wigner functions, $W_{n}(\xi,\pi)$, are given by
\begin{eqnarray}
    W_{n}(\xi,\pi)&=&\frac{(-1)^{n}}{\pi\hbar}e^{-\frac{2H(\xi,\pi)}{\hbar\Omega}}L_{n}\left(\frac{4H(\xi,\pi)}{\hbar\Omega}\right) \nonumber \\
    &=& \frac{(-1)^{n}}{\pi\hbar}e^{-\frac{2I}{\hbar\Omega}}L_{n}\left(\frac{4I}{\hbar\Omega}\right).
\end{eqnarray}
Here $L_{n}\left(\frac{4I}{\hbar\Omega}\right)$ denotes the standard Laguerre polynomials of order $n$, evaluated on the Lewis-Riesenfeld invariant (\ref{I}). The Wigner functions $W_{n}(\xi,\pi)$ agree with those derived exclusively using the wave function method for the time-dependent harmonic oscillator \cite{Koh}. It is worth noting that the Wigner functions in the original $q$-coordinate representation are expressed here in terms of the invariant $I$, rather than the Hamiltonian, as in the case for the simple harmonic oscillator \cite{Curtright}, and thus, the Lewis-Resienfeld invariant is quantized by the relation $I_n=\hbar\Omega (n + 1/2)$ with respect to the time variable $\tau$. This behavior supports the idea that Wigner functions of dynamical systems are parameterized not only by the energy but also by other constants of motion, as highlighted in \cite{Dahl,Belmonte,Braunss}. Moreover, by incorporating the time variable $\tau(t)$, through the relation (\ref{dtau}),
\begin{equation}\label{tau}
    \tau(t)=\int \frac{1}{m(t)\rho^{2}}dt,
\end{equation}
the exponential term in the Fourier-Dirichlet expansion (\ref{FD}), takes the form
\begin{equation}
    e^{-i\tau\Omega(n+1/2)}=e^{-i\Omega(n+1/2)\int \frac{1}{m(t)\rho^{2}}dt}.
\end{equation}
This expression coincides with the time-dependent phase functions obtained by solving the Schr\"odinger equation, employing a complete set of orthonormal eigenfunctions associated with the Lewis-Riesenfeld invariant. These phase functions are essential for deriving the exact general solution of the time-dependent Schr\"odinger equation and are directly related with the energy spectrum \cite{Pedrosa2}. Since the Lagrangian defined in (\ref{Lag}) is time-dependent, the star exponential (\ref{StarEho}), expressed in terms of the original $q$-coordinate representation and the time variable $t$, properly speaking, corresponds to the symbol associated with the time-ordered evolution operator, which is commonly written using Dyson’s series \cite{Takhtajan}. 

\section{The Caldirola-Kanai Oscillator}
\label{sec:CK}

In this section, using of the star exponential of a time-dependent harmonic oscillator derived in (\ref{StarEho}), we determine the time-dependent phase functions and the Wigner functions of the quantum damped harmonic oscillator, commonly known as the Caldirola-Kanai oscillator \cite{Hasse, Greenberger}. The Caldirola-Kanai oscillator is obtained by assuming $\gamma(t)=\gamma_0$, $\omega(t)=\omega_0$, $\Omega=1$ and $m(t)=m_0e^{\gamma_0 t}$, where $\gamma_{0}$, $\omega_{0}$ and $m_{0}$ are constants. Then, the solution of the auxiliary equation (\ref{rho}) is given by \cite{Pedrosa},
\begin{equation}\label{rhoCK}
    \rho(t)=\frac{e^{-\gamma_0 t/2}}{(m_{0}\Omega_{0})^{1/2}}.
\end{equation}
where $\Omega^{2}_{0}=\omega_0^2-\frac{\gamma_0^2}{4}$. Furthermore, the time variable $\tau(t)$ for the Caldirola-Kanai oscillator, as defined in equation (\ref{tau}), takes the form 
\begin{equation}\label{tauCK}
    \tau(t)=\Omega_{0}t.
\end{equation}
The Hamiltonian of the system follows from Eq. (\ref{Ham}),
\begin{equation}\label{HCK}
H_{\mathrm{CK}}(x,p,t)=\left(\frac{p^{2}}{2m_{0}}\right)e^{-\gamma_{0}t}+\frac{1}{2}m_{0}\omega_{0}^{2}e^{\gamma_{0}t}x^{2},
\end{equation}
where the canonical momentum is given by
\begin{equation}
    p=m_{0}e^{\gamma_{0}t}\dot{x}.
\end{equation}
Additionally, the Hamiltonian in the $\xi$-coordinate system is expressed as follows
\begin{eqnarray}\label{ICK}
    H_{\mathrm{CK}}(\xi,\pi)&=&\frac{1}{2}(\pi^{2}+\xi^{2}) \nonumber \\
    &=& \frac{1}{2\Omega_{0}}\left[\frac{p^{2}}{m_{0}}e^{-\gamma_{0}t}+\gamma_{0}px+\frac{m_{0}\gamma_{0}^{2}}{4\Omega_{0}}e^{\gamma_{0}t}x^{2}\right]\equiv I_{\mathrm{CK}}.
\end{eqnarray}
The quantity defined as $I_{\mathrm{CK}}$ corresponds to the Lewis-Riesenfeld invariant for the Caldirola-Kanai oscillator, and satisfies $d I_{\mathrm{CK}}/dt=0$.
By substituting Eqs. (\ref{tauCK}) and (\ref{ICK}) into the star exponential (\ref{StarEho}), we have
\begin{equation}\label{SECK}
        \Exp\left(-\frac{i}{\hbar}\Omega_{0}t H_{\mathrm{CK}}(\xi,\pi)\right)=\cos{\left[\frac{\Omega_{0}t}{2}\right]}\exp\left[\frac{2H_{\mathrm{CK}}(\xi,\pi)}{i\hbar}\tan\left(\frac{\Omega_{0}t}{2}\right)\right].
\end{equation}
 The star exponential function of the Caldirola-Kanai oscillator admits a Fourier-Dirichlet expansion of the form
\begin{equation}\label{FDCK}
    \Exp\left(-\frac{i}{\hbar}\Omega_{0}t H_{\mathrm{CK}}(\xi,\pi)\right)=\sum_{n=0}^{\infty}W_{n}^{\mathrm{CK}}(\xi,\pi)e^{-i\Omega_{0}t(n+1/2)},
\end{equation}
where the diagonal Wigner functions, $W_{n}^{\mathrm{CK}}(\xi,\pi)$, are given by
\begin{equation}
     W_{n}^{\mathrm{CK}}(\xi,\pi)=\frac{(-1)^n}{\pi\hbar}e^{-\frac{2}{\hbar}I_{\mathrm{CK}}}L_n\left(\frac{4I_{\mathrm{CK}}}{\hbar}\right).
\end{equation}
The exponential term, $e^{-i\Omega_{0}t(n+1/2)}$, in the Fourier-Dirichlet expansion (\ref{FDCK}) corresponds to the phase space functions determined in \cite{Pedrosa}, where the exact wave functions for the Caldirola-Kanai oscillator were obtained using the Lewis-Riesenfeld method in the Schr\"odinger picture approach.

\section{Time-dependent frequency harmonic oscillator}
\label{sec:TD-Freq}
In this section, we obtain the time-dependent phase functions and the Wigner functions of the time-dependent frequency harmonic oscillator. Many physical systems exhibit behaviors that can be modeled by a time-dependent frequency harmonic oscillator. Examples include ions oscillating in one-dimensional Pauli traps \cite{Paul}, radiation fields propagating through time-dependent dielectric media \cite{dielectric}, among others. A time-dependent harmonic oscillator is obtained by assuming $m(t)=1$, $\gamma(t)=0$ and $\Omega=1$. According to equation (\ref{Ham}), the Hamiltonian of the system reads
\begin{equation}
    H_{\mathrm{TDF}}(x,p,t)=\frac{1}{2}p^{2}+\frac{1}{2}\omega(t)^{2}x^{2}.
\end{equation}
For this dynamical system, the Lewis-Riesenfeld invariant takes the form \cite{LR, Moya},
\begin{equation}
    I_{\mathrm{TDF}}=\frac{1}{2}(p\rho-x\dot{\rho})^{2}+\frac{x^{2}}{\rho^{2}},
\end{equation}
where $p=\dot{x}$ corresponds to the canonical momentum and $\rho$ satisfies the auxiliary equation (\ref{rho}), given by the Ermakov-Pinney nonlinear differential equation \cite{Pinney},
\begin{equation}
    \ddot{\rho}+\omega^{2}(t)\rho=\frac{1}{\rho^{3}}.
\end{equation}
The general solution of this nonlinear differential equation is given by
\begin{equation}
    \rho(t)=\left(u^{2}(t)+\frac{1}{W(u,v)^{2}}v^{2}(t)\right)^{1/2},
\end{equation}
where $u(t)$ and $v(t)$ form a fundamental set of solutions of the linear equation
\begin{equation}
    \ddot{\rho}+\omega^{2}(t)\rho=0,
\end{equation}
and $W(u,v)$ denotes the Wronskian associated to the fundamental solutions. For this case, the star exponential function (\ref{StarEho}) acquires the form
\begin{equation}\label{STTDF}
    \Exp\left(-\frac{i}{\hbar}\tau H_{\mathrm{TDF}}(\xi,\pi)\right)=\sum_{n}^{\infty}W_{n}^{\mathrm{TDF}}(\xi,\pi)e^{-i\tau(n+1/2)},
\end{equation}
where the diagonal Wigner functions, $W^{\mathrm{TDF}}_{n}(\xi,\pi)$, are given by
\begin{equation}
    W^{\mathrm{TDF}}_{n}(\xi,\pi)=\frac{(-1)^{n}}{\pi\hbar}e^{-\frac{2I_{\mathrm{TDF}}}{\hbar}}L_{n}\left(\frac{4I_{\mathrm{TDF}}}{\hbar}\right).
\end{equation}
Moreover, by introducing the time variable $\tau$, through the relation (\ref{tau}) 
\begin{equation}
    \tau(t)=\int\frac{1}{\rho^{2}}dt,
\end{equation}
the exponential term in the Fourier-Dirichlet expansion (\ref{STTDF}), takes the form
\begin{equation}
    e^{-i\tau(n+1/2)}=e^{-i(n+1/2)\int \frac{1}{\rho^{2}}dt}.
\end{equation}
This expression corresponds to the phase space functions determined in \cite{Moya2}, where the exact wave functions for the time-dependent frequency harmonic oscillator were obtained by means of the Koopman-von Neumann (KvN) approach to mechanics.

\section{Conclusions}
\label{sec:conclu}

We have addressed the construction of the Wigner distribution and the star exponential function for the mass and frequency time-dependent harmonic oscillator.  Such structures are fundamental within the Deformation quantization program as they encode the probability density and the time evolution for a quantum mechanical system, respectively.   In particular, we have explored a recent proposal to construct the star exponential function by suitably Fourier transforming the quantum mechanical propagator.  In the case of our interest here, the star exponential function was straightforwardly obtained by implementing a time-depending scaling and a non-linear time variable that convert the time-dependent Lagrangian into a simple harmonic oscillator with constant frequency.   In consequence, the Hamiltonian function for the latter system may be directly identified as the Lewis-Riesenfeld invariant, for which the star exponential function was explicitly built.  Further, invoking the Fourier-Dirichlet expansion of the star exponential, we were able to construct the diagonal Wigner distribution that, in turn, imposed a quantization on the eigenvalues of the Lewis-Riesenfeld invariant corresponding to those of conserved energy in the scaled coordinate system.     Further, the exponential term in the Fourier-Dirichlet expansion was shown to be coincident with the time-dependent phase functions obtained within the standard Schrödinger formalism for the time-dependent harmonic oscillator.   In addition, we have explored our previous results to encounter the star exponential and the time-dependent phases for the Caldirola-Kanai and the time-dependent frequency harmonic oscillators.   All of our results for these last models were in agreement with previous results developed within different quantum schemes.

In conclusion, we would like to emphasize that the relation between the star exponential and the propagator has been effectively used to explore the time-dependent model of our interest here, thus paving the way for the analysis of physical phenomena with oscillatory damping or time-varying potentials, including possible extensions to their field theoretical counterparts.

\section*{Acknowledgments}

The authors would like to acknowledge support from SNII SECIHTI-Mexico. JBM acknowledge financial support from Marcos Moshinsky foundation and thanks the Dipartimento di Fisica ``Ettore Pancini" for the kind invitation and its generous hospitality.  AM acknowledges financial support from COPOCYT under project 2467 HCDC/2024/SE-02/16 (Convocatoria 2024-03, Fideicomiso 23871). MSC acknowledges SECIHTI-Mexico for supporting the postdoctoral fellowship.

\section*{References}

\bibliographystyle{unsrt}

\end{document}